\documentclass[useAMS,usenatbib]{mn2e}
\usepackage{epsfig}
\usepackage{graphicx}
\usepackage{aas_macros}
\usepackage{color}
\usepackage{amsmath}
\usepackage{amssymb}
\usepackage{epsfig}
\usepackage{natbib}

\newcommand{\oversim}[2]{\protect{\mbox{\lower0.5ex\vbox{%
   \baselineskip=0pt\lineskip=0.2ex
   \ialign{$\mathsurround=0pt #1\hfil##\hfil$\crcr#2\crcr\sim\crcr}}}}} 
\newcommand{\simgreat}{\mbox{$\,\mathrel{\mathpalette\oversim>}\,$}} 
\newcommand{\simless} {\mbox{$\,\mathrel{\mathpalette\oversim<}\,$}} 

\begin{document}

\title[Binary stars as probes of dark substructures in dSphs]{Binary stars as probes of dark substructures in dwarf galaxies}   
   
\author[Pe\~{n}arrubia et al.]{Jorge Pe\~{n}arrubia$^{1}$\thanks{jorpega@ast.cam.ac.uk}, Sergey E. Koposov$^{1,2}$, Matthew G. Walker$^{1}$, Gerry Gilmore$^{1}$, 
\newauthor  
N. Wyn Evans$^{1}$ \& Craig D. Mackay$^{1}$\\
$^{1}$Institute of Astronomy \& Kavli Institute for Cosmology, University of Cambridge, Madingley Road, Cambridge, CB3 0HA, UK \\
$^{2}$Sternberg Astronomical Institute, Universitetskiy pr. 13, 119992 Moscow, Russia
}
\maketitle  

\begin{abstract}   
We use analytical and N-body methods to examine the survival of wide stellar binaries against repeated encounters with dark substructures orbiting in the dark matter haloes of dwarf spheroidal galaxies (dSphs). Our models adopt cosmologically-motivated conditions wherein dSphs are dark-matter dominated systems that form hierarchically and orbit about a host galaxy. Our analytical estimates show that wide binaries are disrupted at a rate that is proportional to the local density of dark substructures averaged over the life-time of the binary population.  
The fact that external tides can efficiently strip dark substructures from the outskirts of dSphs implies that the present number and distribution of binaries is strongly coupled with the mass evolution of individual galaxies. Yet we show that for the range of dynamical masses and Galactocentric distances spanned by Milky Way dSphs, a truncation in the separation function at $a_{\rm max}\simless 0.1 $pc is expected in {\it all} these galaxies. An exception may be the Sagittarius dSph, which has lost most of is dark matter envelope to tides and is close to full disruption. Our simulations indicate that at separations larger than $a_{\rm max}$ the perturbed binary distribution scales as $dN/da\propto a^{-2.1}$ independently of the mass and density of substructures. These results may be used to determine whether the binary separation function found in dwarf galaxies is compatible with the scale-free hierarchical picture that envisions the existence of dark substructures in all galactic haloes. We show that the ACS camera on board of the Hubble telescope may be able to test this prediction in dSphs at heliocentric distances $\simless 100$ kpc, even if the binary fraction amounts only 10\% of the stellar population.

\end{abstract}   
\section{Introduction} \label{sec:intro}   
Perhaps one of the most striking predictions from the present cosmological paradigm (Cold Dark Matter, CDM) is the existence of dark matter haloes devoid of baryonic material, the so-called ``dark galaxies''. Such systems arise because primordial gas is unable to cool efficiently by collisional exicitation of $H_2$ molecules (Haiman et al. 2000), whose formation is limited to haloes with virial temperatures above $10^4 $K (i.e. virial masses $\simgreat 10^8 M_\odot$). Haloes with masses below $\sim 10^8 M_\odot$ are thus expected to be dark.
The fact that the CDM halo mass function diverges as $dN/d m\propto m^{-1.9}$ (Diemand et al. 2008; Springel et al. 2008 and references therein) implies that dark galaxies dominate in number over luminous ones. Indeed, we expect of the order of $10^{12}$ subhaloes with masses within $10^{-6}$--$10^8 M_\odot$ lingering in a Milky Way-size halo (Diemand et al. 2005).

How can we possibly test the existence of dark galaxies? If dark matter is in the form of a neutral relic particle once in thermal equilibrium in the early universe, these objects may be lit up by pair annihilation that would result in gamma-ray emmissions. Unfortunately, testing this scenario on the mass scale of dark galaxies is well beyond the present technological capabilities for the vast majority of particle candidates (e.g. Diemand et al. 2007). The second alternative is to detect dark haloes via gravitational perturbations of visible objects.

In this regard very wide binaries, i.e. those with separations $a\simgreat 100$ AU, may enable simple experiments to test for the presence of dark substructures, for even extremely weak tidal perturbations can disrupt them. In the stellar halo of the Milky Way (hereafter MW), perturbers can be identified as inhomogeneities in the Galactic potential. Some examples are stellar clusters, molecular clouds and compact objects (i.e. MACHOS), which altogether cover a mass range that overlaps with that of dark galaxies (Carr 
\& Sakellariadou 1999, Chanam{\'e} \& Gould 2004). Hence isolating the effects of different pertubers on binaries is not straightforward.

Much more interesting targets are binaries in {\it dwarf galaxies}. The reasons are several: (i) dwarf spheroidal galaxies (hereafter dSphs) have the largest dark matter content of any known galaxy type in the Universe (e.g. Mateo 1998, Gilmore et al. 2007). In dSphs the fraction of dark to baryonic mass is so large ($10^1$--$10^3$) that the presence of baryonic inhomogeneities in the dwarf potential can be safely neglected. (ii) As a result of their hierarchical formation, dwarf galaxies orbiting the MW are expected to contain bound substructures. These objects are typically referred to as (sub-)subhaloes (hereafter sub$^2$haloes) in cosmological simulations. Given that the estimated virial mass of the MW dSphs is of the order of $M_{\rm vir}\sim 10^9 M_\odot$ (e.g. Pe\~narrubia et al. 2008a, Walker et al. 2009) and that in CDM the halo mass function is scale-free, all dark matter substructures orbiting in dwarfs are expected to fall below the star formation threshold and thus be dark. (iii) Dwarf galaxies are embedded in haloes with estimated dark matter densities that are several orders of magnitude higher than in the solar neighbourhood. In a scale-free cosmology this implies a higher number density of perturbers and thus a higher chance to detect their effects on a pre-existing binary population. (iv) The stellar populations of dSphs are generally dominated by old and intermediate-age components (e.g. Mateo 1998 and references therein), which enhances the cumulative effect of gravitational perturbations.

Unfortunaltey, although binaries with intermediate ($\simless 100$ AU) separations have been studied in stellar clusters with some detail in the past (e.g. White \& Ghez 2001; Patience et al. 2002), little is known about the number and distribution of binary stars in dwarf spheroidal galaxies (e.g. Wyse et al. 2002 estimate that 5\% of the stars in the Ursa Minor dwarf correspond to unresolved binaries from the number of blue stragglers detected in this galaxy).
 
This paper is aimed at studying the disruption of wide ($\simgreat 100$ AU) binaries as a result of interactions with dark substructures orbiting in dwarf galaxies, as well as exploring whether the effects can be detected with our present observational capabilities. The arrangement of the paper is as follows: \S2 outlines the analytical analysis, \S3 tests its consistency through N-body simulations, \S4 applies the result to the Milky Way dwarf galaxy population, \S5 estimates the feasibility of detecting wide binaries in dwarfs and \S6 provides a summary of the results.

\section{Catastrophic vs. Diffusive encounters}
\label{sec:catdif}   
Binary stellar systems travelling through a homogeneous sea of perturbers with mass $M_p$, number density $n_p$ and density $\rho_p=M_pn_p$ suffer the cumulative effect of repeated encounters over their life times. Assuming that the relative velocity of encounters $V_{\rm rel}$ is isotropic and follows a Maxwellian distribution, the change of energy per unit mass averaged over a suite of binary orbits with different orbital phases and eccentricities is (Binney \& Tremaine 2008; eq. 8.48)
\begin{equation}
<\Delta E>=\frac{7G^2M_p^2a^2}{3 V_{\rm rel}^2 b^4}U(b/r_h);
\label{eq:de}
\end{equation}
where $b$ is the impact parameter and $r_h$ the half-mass radius of the perturber. For extended bodies that follow a Hernquist (1990) mass profile $U(x)\approx x^{2.5}/(1+x)^{2.5}$, the point-mass approximation, $U\approx 1$, is recovered at $x\equiv b/r_h\simgreat 1$.
For simplicity we shall hereafter assume that the point-mass approximation can be applied to all perturbers, whilst leaving a more stringent analysis for Appendix~\ref{sec:pmext}.

During the life time of a binary, $<\Delta E>$ can grow (i) as a result of repeated encounters that are not fatal for the system or (ii) through single encounters that completely shatter the system. In the former case the energy change gained after single interactions is small, $<\Delta E>/|E|\ll 1$, where $E=-G M_b/(2a)$ is the binding energy of the binary. This is the so-called {\it diffusive regime}.
Perturbations occur in a {\it catastrophic regime} if $<\Delta E>/|E|\gg 1 $. For point-mass perturbers the condition for catastrophic encounters is
\begin{equation}
b\ll 1.5\bigg[\frac{G M_p^2 a^3 }{M_b V_{\rm rel}^2}\bigg]^{1/4}\equiv b_{\rm min},
\label{eq:bmin}
\end{equation}
so that when a tidal perturbation is sufficiently strong to be fatal for a binary even at $b_{\rm min}/a\gg 1$, we consider that the effects of single, catastrophic interactions dominate over diffusive ones.

To identify the dominant regime of perturbations we define the quantity $M_{p,{\rm min}}$ as the perturber mass that yields $b_{\rm min}/a=1$ in eq.~(\ref{eq:bmin}). Subsequently we scale this equation to parameters typically found in dwarf galaxies. Since stars in these systems show no rotation and a typical velocity dispersion $\sigma_\star\sim 10$ km/s (Mateo 1998; Walker et al 2007) the relative velocity of encounters can be assumed to be of the order of $V_{\rm rel}=\sqrt{2} \sigma_\star$. 
Using $M_b=1 M_\odot$ and $a=0.1$ pc as fiducial binary parameters we find

\begin{equation}
M_{p,{\rm min}}\approx 25 M_\odot \frac{V_{\rm rel}}{\sqrt{2} \cdot 10 {\rm km s}^{-1}}\bigg( \frac{M_b}{1 M_\odot}\frac{a}{0.1 {\rm pc}} \bigg )^{1/2}; 
\label{eq:mpmin}
\end{equation}
so that the boundary between diffusive and catastrophic encounters is placed at  $M_p\sim 10^2 M_\odot$ for the range of separations relevant in this study.

In this work we identify ``perturbers'' with dark matter sub$^2$haloes orbiting in the dark matter haloes of dwarf spheroidal galaxies.
Recent estimates show that the virial masses of the brightest MW dwarfs span a relatively small range of values around $M_{\rm vir}\sim 10^{9} M_\odot$ (e.g. Pe\~narrubia et al. 2008a).   Adopting the subhalo mass function found in CDM simulations, sub$^2$haloes are expected to have masses $M_p\simless 0.1 M_{\rm vir}\approx 10^8 M_\odot \gg M_{p,{\rm min}}$ (Navarro, Frenk \& White 1997; hereafter NFW). Since the expected perturber masses are many orders of magnitude larger than $M_{p,{\rm min}}$ the effect of diffusive interactions can be safely neglected in our study

Indeed, the fact that interactions between dark matter substructures and wide binaries can be treated as catastrophic perturbations is a remarkable result, because in this regime the disruption time scales as (e.g. Binney \& Tremaine 2008; eq. 8.51)
\begin{equation}
t_d\simeq k_{\rm cat}\frac{1}{G \rho_p}\bigg(\frac{G M_b}{a^3}\bigg)^{1/2};
\label{eq:td}
\end{equation}
where $k_{\rm cat}\simeq 0.07$ (Bahcall, Hut \& Tremaine 1985). Note that the disruption of wide binaries is independent of both the relative velocity of the encounter and the pertuber mass. Thus, observed binary distributions in dwarf galaxies can yield significant constraints on the amount and distribution of dark matter substructures, $\rho_p$, in these systems.

In what follows, it is useful to define the quantity
\begin{equation}
a_{\rm max}\equiv \bigg( \frac{k_{\rm cat}}{G \rho_p \Delta t}\bigg)^{2/3} (G M_b)^{1/3};
\label{eq:amax}
\end{equation}
so that binaries with initial separations $a\simless a_{\rm max}$ and a mass $M_b$ are expected to be disrupted over a time-scale $\Delta t$.

\begin{figure}  
\includegraphics[width=84mm]{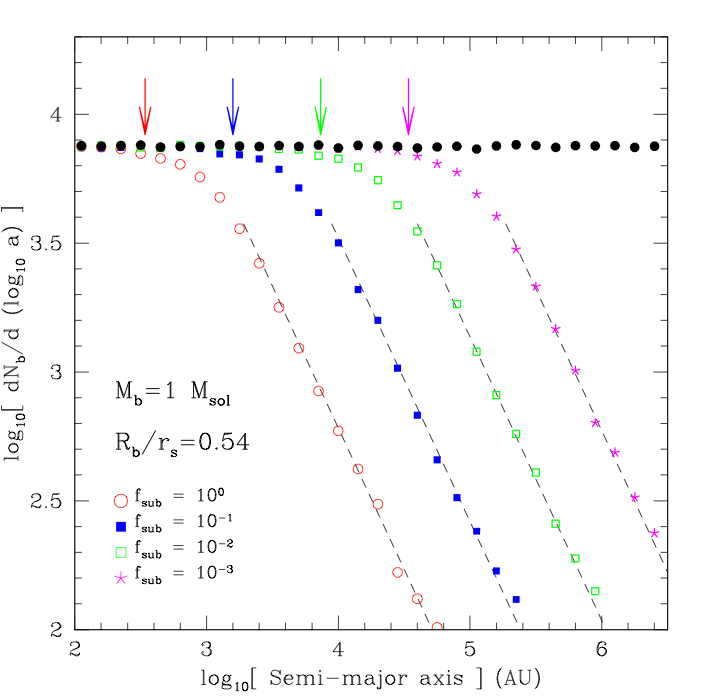}
\caption{Binary distribution as a function of semi-major axis. $N_b=3\cdot 10^5$ binaries are generated following an arbitrarily chosen flat distribution represented as filled circles. Binary and perturber orbits are drawn from an NFW halo N-body model with $M_{\rm vir}=10^9 M_\odot$ and $r_s=1.11$ kpc evolved during 14 Gyr in isolation (see text). All binary particles move on orbits with apocentres smaller than $R_b$. 
The perturber mass spectrum scales as $dN(M_p)/dM_p \propto M_p^{-1.9}$, and the total mass in perturbers is $f_{\rm sub}\equiv \sum_{i=1,...,N_p} M_{p,i} / M_{\rm vir}$. For these particular models we use $N_p=10^3$, although these results are independent of our choice of $N_p$ and $N_b$. Arrows show the location of $a_{\rm max}$ as derived from eq.~(\ref{eq:amax}). Dashed lines show that for $a\gg a_{\rm max}$ the separation distribution scales $dN_b/da\propto a^{-2.1}$. }  
\label{fig:dist}  
\end{figure}

\section{Monte-Carlo simulations}
\label{sec:test} 
To test the consistency of the analytical predictions we simulate the effects of encounters between dark matter subhaloes and binaries using a standard Monte-Carlo technique.  We adopt an ``unperturbed'' initial separation distribution of binaries that follows \"Opik's (1924) law $dN_b/da\propto a^{-1}$ within $10^1$--$10^{7}$ AU or, equivalently, that is uniform over the interval $1.0<\log_{10}(a/{\rm AU})<7$. For simplicity, we assume $U=1$ and a fiducial binary mass $M_b=1 M_\odot$ in eq.~(\ref{eq:de}) (see Appendix~\ref{sec:pmext} for simulations where perturbers are treated as extended objects).

The orbital distributions of ``stars'' and  ``perturbers'' are drawn from the N-body models of Pe\~narrubia et al. (2008b) evolved in isolation for a Hubble time, so that $\Delta t=14$ Gyr in eq.~(\ref{eq:amax}). These models follow an NFW profile with a virial mass and concentration $M_{\rm vir}=10^9 M_\odot$ and $c_{\rm vir}=23.1$, respectively. Using the concordance cosmological parameters (e.g. Spergel et al. 2007) this choice corresponds to scale and virial radii $r_s=1.11$ kpc and $r_{\rm vir}=25.6$ kpc, respectively.

From the N-body models, $N_p$ perturber particles are randomly chosen. We also select $N_b$ binary particles with apocentric distances to the dwarf centre smaller than $R_b$, which accounts for the fact that stars are spatially segregated within the dark matter haloes of dSphs (Pe\~narrubia et al. 2008a). Given that the typical half-light radii of bright ($M_V<-8$) dSphs is $\sim 300$ pc (Mateo 1998), we choose a fiducial $R_b=600$ pc, i.e. $R_b/r_s=0.54$.  Subsequently, we follow for a Hubble time the relative position between the star ``i'' and the perturbers, ${\bf r}_{i,j}={\bf r}_{b,i}-{\bf r}_{p,j}$, where $j=1,...,N_p$ and $i=1,...,N_b$. Whenever the quantity $|{\bf r}_{i,j}|\equiv b$ finds a minimum we store the parameters of the encounter $(b,V_{\rm rel})_{i,j}$. 

Perturber masses are calculated adopting the subhalo mass function typically found in CDM N-body simulations, which scales as $dN(M_p)/dM_p \propto M_p^{-\alpha}$, where $\alpha\approx 1.9$ (Springel et al. 2008). The fact that $\alpha<2.0$ is crucial because, although the number of perturbers diverges at low mass, the total mass in perturbers is dominated by a few massive substructures with $M_p\simless 0.1 M_{\rm vir}$. Here we derive the normalization of the perturber mass spectrum by defining the quantity $f_{\rm sub}\equiv \sum_{i=1,...,N_p} M_{p,i} / M_{\rm vir}$.

Fig.~\ref{fig:dist} shows that dark matter substructures efficiently destroy binaries with long semi-major axes, and that this process clearly strengthens as the mass in dark substructures, indicated by $f_{\rm sub}$, increases. The result is a final separation distribution that is {\it truncated} with respect to the initial profile. Remarkably, the perturbed part of the distribution can be accurately approximated by $dN_b/d a \propto a^{-2.1}$ for $a\simgreat a_{\rm max}$, {\it independently of perturber mass and density}\footnote{Binary populations experiencing perturbations in a diffusive regime also show power-law distributions. However, the slope is not unique and depends on several encounter parameters (Yoo, Chanam\'e \& Gould 2004)}. 

For each simulation we mark the value of $a_{\rm max}$ derived from eq.~(\ref{eq:amax}) with arrows. We define the (local) density of perturbers as 
\begin{equation}
\langle \rho_p \rangle=\frac{f_{\rm sub} M_{\rm vir}}{ 4\pi/3 R_b^3}.
\label{eq:rhopl}
\end{equation}
Interestingly, the analytical value of $a_{\rm max}$ accurately signals the binary separation at which deviations from the unperturbed distribution start to become obvious (see also Fig.~\ref{fig:pmext}). 

In the following Section we apply these results to Milky Way dSphs in order to examine whether their binary populations can probe the existence of dark substructures orbiting within their dark matter haloes.

\begin{figure}
\includegraphics[width=84mm]{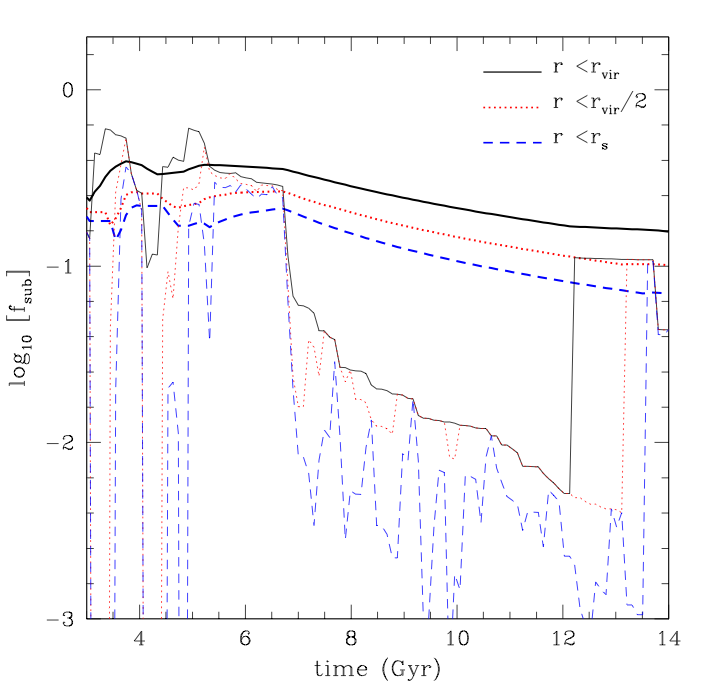}
\caption{Evolution of the subhalo mass fraction $f_{\rm sub}\equiv \sum M_{\rm sub}/M_{\rm vir}$ measured at different distances from the centre of a Milky Way-size halo, i.e. $M_{\rm vir}=10^{12}M_\odot$ (realization taken from Pe\~narrubia et al. 2010). The infall of massive substructures can be easily identified as {\it jumps} in the value of $f_{\rm sub}$. Although this quantity suffers strong variations, the time-average values (thick lines) remain fairly constant in time.  }
\label{fig:fsub}
\end{figure}

\section{Binary disruption in Milky Way dSphs}
\label{sec:dSphs}

Estimating the time-averaged density of substructures in dwarf galaxies orbiting about a host is not exempt from complexities.
In {\it field} haloes the process of accretion, tidal stripping and merger of small substructures proceeds uninterruptedly in a way that keeps the time-averaged local subhalo density practically constant in time
This is illustrated in Fig.~\ref{fig:fsub}, where we show the evolution of the subhalo mass fraction $f_{\rm sub}(<r)\equiv \sum M_{\rm sub}(<r)/M_{\rm vir}$ calculated at different radii from a Milky Way-size halo realization\footnote{Since in $\Lambda$CDM the hierarchical formation of haloes is scale-free, we expect a similar evolution also on the mass scale of dSphs.}. Infall of massive substructures can be easily identified as ``jumps'' in the value of $f_{\rm sub}(<r_{\rm vir})$ that propagate towards smaller radii as massive substructures sink into the inner-most regions of the host through dynamical friction (thin lines). Remarkably, although the local mass in substructures suffers strong variations, the time-averaged value of $f_{\rm sub}$ (thick lines) stays relatively constant at all radii.

 However, this process is interrupted when field galaxies are accreted into a larger host and become {\it satellites}: On the one hand, the accretion of external sub$^2$haloes stops after satellite galaxies cross the virial radius of the parent galaxy. Hence, substructures that are tidally disrupted stop being replenished. On the other hand, a fraction of the existing substructures is lost after each pericentric passage  through tidal stripping. The combination of both effects tend to lower the mass fraction of sub$^2$haloes in satellites in a monotonic fashion and induce a correlation between the present value of $f_{\rm sub}$ and the location of satellites within the main halo.

\begin{figure}  
\includegraphics[width=84mm]{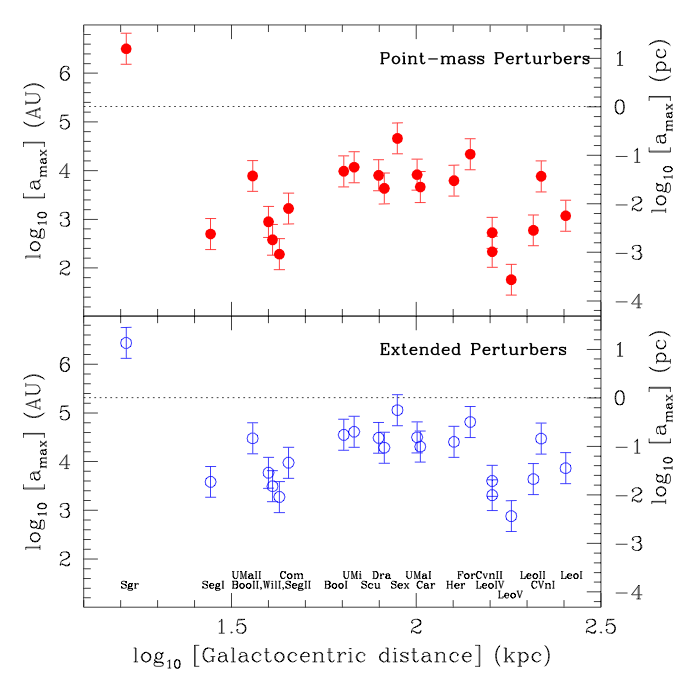}
\caption{Expected truncation in the separation function $a_{\rm max}$ of a binary population with $M_b=1 M_\odot$ that was formed at $t_b=2$ Gyr after the Big Bang (see text). {\it Upper} and {\it lower} panels assume that dark structures can be treated as point-mass and extended (see Appendix~\ref{sec:pmext}) perturbers, respectively. The horizonthal dotted lines marks a separation of 1 pc, which roughly corresponds to the largest separation expected in a binary sample. }  
\label{fig:dsphs}  
\end{figure}

Dealing with these aspects of the problem in a consistent way requires the aid of cosmological N-body simulations with a spatial and time resolution that, unfortunately, go beyond the present numerical capabilities.
Here we make three crude assumptions that allow us to estimate the location of the truncation in the binary separation function of MW Sphs in a simple way. First we use eq.~(\ref{eq:rhopl}) to estimate the time-averaged local density of perturbers. Second, since the amount of stripping experienced by the MW dSphs is difficult to gauge from observational data (Pe\~narrubia et al. 2009), we use the statistical correlation between $f_{\rm sub}$ and Galactocentric distance found in the Aquarius run (Springel et al. 2008) to guide our estimates\footnote{Note that the Aquarius models do not include baryons.  Since discs strongly enhance the tidal stripping of satellite galaxies (D'Onghia et al. 2010,  Pe\~narrubia et al. 2010), this correlation may be stronger in spiral galaxies.}, which can be expressed as $f_{\rm sub}(R)=f_0(R/R_{\rm vir})^{\nu}$ for $R\le R_{\rm vir}$, where $R$ is the Galactocentric distance of the dwarf, $f_0\simeq 0.2$ and $\nu\simeq 2.0$, with a log normal scatter of $\Delta \log_{10}\sigma =0.5$ at fixed radii. Third, we adopt Walker et al. (2009) results, who show that the estimated dynamical masses of all MW dSphs are compatible with these systems being embedded in dark matter haloes with $M_{\rm vir}\sim 10^9 M_\odot$. 

Under these assumptions, the expected value of $a_{\rm max}$ for binaries with $M_b=1 M_\odot$ is shown in Fig.~\ref{fig:dsphs} assuming that dark substructures can be treated as point-mass ({\it upper panel}) and extended ({\it lower panel}, see Appendix~\ref{sec:pmext}) perturbers acting on the binary population. Error bars account for the scatter found in the cosmological relation between $f_{\rm sub}$ and Galactocentric distance. For ease of reference, we add a horizonthal dotted line to mark a separation of 1 pc, which roughly corresponds to the maximum separation found in (spectroscopically confirmed) binaries in the stellar halo of the MW (Quinn et al. 2009). 

This Figure clearly shows that {\it all} dwarf galaxies in the MW are expected to have truncated binary separation functions within the context of cosmologically-motivated CDM haloes. Focusing on the extended perturber models, truncation may be visible at separations as small as $\sim 10^{-2}$ pc for most of the ultra-faint dSphs (Belokurov et al. 2007, 2009), which have stellar sizes $R_b=2\times r_{\rm half}\simless 100$pc. ``Classical'' dwarfs on the other hand are more extended, $500 \simless (R_b/pc)\simless 800$, and tend to reside at further Galactocentric distances ($\simgreat 80$ kpc). For these systems our estimates suggest that the binary population should have a separation function truncated beyond $\simless 0.1$ pc. A notable exception however is the Sagittarius (Sgr) dwarf, which is currently close to full tidal disruption (Niederste-Ostholt et al. 2010). Putting its Galactocentric distance in the Aquarius relation yields a very low content of substructures, $f_{\rm sub}\simeq 2\cdot 10^{-4}$, which likely implies an unperturbed ($a_{\rm max}\simgreat 1$ pc) binary function.

\begin{figure}  
\includegraphics[width=84mm]{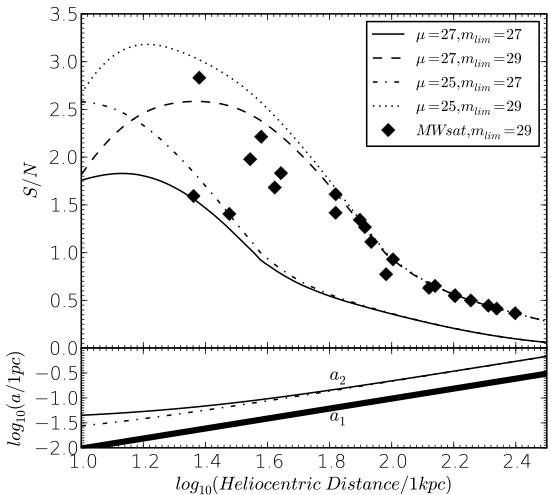}
\caption{ {\it Upper panel:} Effective signal to noise ratio of binaries with respect to random pairs $S/N=N_{\rm bin}/{\sqrt{N_{\rm bin}+N_{\rm ran}}}$ in dSphs with different surface brightnesses ($\mu$) as a function of heliocentric distance. We adopt the angular resolution and (a single) field of view of the ACS camera on board of the HST. {\it Lower panel:} Interval of separations $(a_1,a_2)$ that maximizes the $S/N$ ratio. Note that $a_1$ is simply the telescope resolution times heliocentric distance. }  
\label{fig:sn}  
\end{figure}  
\section{Observational prospects}
\label{sec:obsteo} 
The analysis of the preceding Section suggests that the separation of binaries in dSphs may scale as $dN/da\propto a^{-2.1}$ for $a\simgreat 0.1 $pc.  Here we examine whether observations of large separation binary populations in Milky Way dSphs are feasible with existing instrumentation.

Let the total number of stars in an observed part of the sky be
\begin{equation}
N_{\infty}=\Omega\,\mu\,N_t,
\end{equation}
where $\Omega$ is the survey area, $\mu$ is the surface brightness of the dwarf galaxy and $N_t=\int_{-\infty}^{\infty} f(M)\,dM$ is the total number of stars per solar luminosity in a dwarf galaxy with distance modulus $(m-M)$ and stellar luminosity function $f(M)$. 

The number of stars which are actually detected (i.e. above the
detection limit) is
\begin{equation}
N_{obs}=\Omega\,\mu\,N_t\,g(\Delta m),
\end{equation}
where $g[\Delta m]=\int_{-\infty}^{\Delta m} f(M)\,dM / N_t$, with $\Delta m\equiv m_{lim}-(m-M)$, denotes the fraction of stars above the limiting magnitude $m_{lim}$.

Thus the number of physical binaries observable in a survey ($N_{\rm bin}$) can be written as
\begin{equation}
N_{\rm bin}=g(\Delta m) N_{obs}\int\limits_{a_1}^{a_2}P(a)\, da =\Omega\, \mu\, N_t g^2[\Delta m]
\int\limits_{a_1}^{a_2}P(a)\, da ;
\label{eq:nbin}
\end{equation}
where $P(a)\,da$ is the fraction of binaries with separations in the interval $a$ and $a+da$; $a_1$ is the minimal separation we are able to resolve in our survey and $a_2$ is the maximal separation.

The observability of wide binaries depends on the contamination by random
stellar pairs. Assuming that
the density of the stars in the dwarf galaxy is significantly higher than the
density of foreground/background object, the number of random pairs with the range of separations $a_1<a<a_2$ can be easily estimated from the surface density of observed stars $N_{obs}/\Omega$ and the angular area enclosed within the annuli $a_1/D$ and $a_2/D$

\begin{equation}
N_{\rm ran} = \pi \Omega \mu^2 N_t^2g^2[\Delta m]\frac{(a_{2}^2-a_{1}^2)}{D^2};
\label{eq:nran}
\end{equation}
where $D$ is the distance to the galaxy. 

The detectability of binaries in a given separation range is determined by an
effective signal to noise ratio $S/N=N_{\rm bin}/{\sqrt{N_{\rm bin}+N_{\rm ran}}}$. And for a given survey the minimal separation $a_1$ is determined by the PSF width, whilst the maximum
separation $a_2$ should be chosen to maximize $S/N$.

We estimate the $S/N$ expected from one exposure at the Advanced
Camera for Surveys (ACS) on board of the Hubble Space Telescope (HST). We assume that the
separations are distributed according to \"Opik's law (1924), i.e. $P(a)\propto a^{-1}$, and that the fraction of wide binaries with separations $10^{-3}<(a/1{\rm pc}) <1$ is 10\%, i.e. similar to that found in the stellar halo of the Milky Way  (Longhitano \& Binggeli 2010). The stellar
luminosity functions were taken from Dartmouth Stellar Evolution Program (Dotter
et al. 2008) assuming the Chabrier mass function (Chabrier 2001) with
$M_c=0.08\,M_\odot$ and $\sigma=0.952$ (Bastian et al. 2010)
and old (12~Gyr) metal-poor ([Fe/H]=-2) stellar populations. Given the PSF
width of ACS (Jee et al. 2007), we use a minimal separation of
4x0.05'' pixels.

Figure~\ref{fig:sn} shows the resulting signal-to-noise for dwarf galaxies at different
heliocentric distances.  Diamonds show the expected $S/N$ value for MW
dwarf satellites adopting $m_{lim}=29^m$. The bottom panel shows the range of binary
separations probed by HST as a function of distance, where $a_1$ is set by resolution and $a_2$ is selected to maximize the S/N ratio. 

This Figure shows two interesting points. First, the surface
brightness of the dwarf is not the driving parameter for the detection of binaries.
Instead the most relevant quantities are the distance to the dwarf and
the depth of the observations. Second, the range of binary separations probed by
HST seems to be reasonable to test the predictions of our paper, but in order to
put significant limits on the binary fraction in dwarfs, several deep exposures
are needed. Note that Figure~\ref{fig:sn} shows the results of a single ACS exposure. For 10 deep
exposures the detection signal would boost to 5.8, 5.0, 4.5, 4.4, 4.2 and 4.0 for Coma, Ursa Minor, Bootes I, Ursa Major II,  Sculptor and Draco, respectively.

\section{Summary}
The present cosmological paradigm, $\Lambda$CDM, predicts the presence of dark matter substructures devoided of baryonic material (the so-called ``dark galaxies'') in all galactic haloes on all mass scales. 
We have considered in this paper the possibility of using wide stellar binaries to uncover the existence of these objects in dwarf spheroidal galaxies (dSphs), whose estimated dark matter densities surpass those of any known galaxy in the Universe.

Our analytical estimates suggest that a large fraction of wide binaries can be wiped out due to tidal encounters with dark substructures. These interactions occur in a ''catastrophic'' regime, wherein the disruption rate is proportional to the density of perturbers averaged over the life-time of the binary population $\langle \rho_p \rangle$. 

We find that in this regime the perturbed binary distribution has a separation function that scales as $dN/da\propto a^{-2.1}$ at $a\simgreat a_{\rm max}$, where $a_{\rm max}\propto \langle \rho_p \rangle^{-2/3}$. 
Our estimates show that for the range of dynamical masses and Galactocentric distances spanned by Milky Way dSphs, {\it all} dwarfs are expected to have a truncation in the separation function at $a_{\rm max}\simless 0.1 $pc within the CDM context.

We examine in which dSphs observations of large separation binaries are feasible with existing instruments, finding that the ACS camera on board of the Hubble telescope would be able to test the predictions enclosed in this paper in dSphs that locate at heliocentric distances $\simless 100$ kpc, even if the binary fraction amounts only 10\% of the stellar population. 

Altogether these considerations may pose a strong test through which $\Lambda$CDM shall soon have to pass.

{}

\appendix

\section{Point-mass vs. extended perturbers}\label{sec:pmext}
\begin{figure}  
\includegraphics[width=84mm]{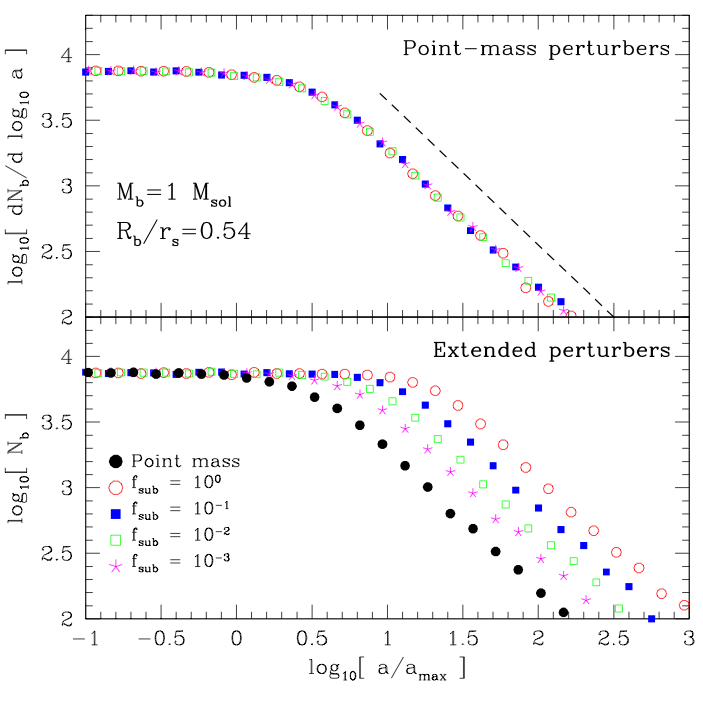}
\caption{Binary separation distribution in simulations where dark structures are treated as point masses ({\it upper panel}, see Fig.~\ref{fig:dist}) and extended objects ({\it lower panel}). In both panels the binary semi-major axis is normalized to the binary maximum separation $a_{\rm max}$, as computed from eq.~(\ref{eq:amax}). 
Note that the perturbed binary population scales as $dN_b/d a \propto a^{-2.1}$ regardless of the inner structure of the perturber objects (dashed line). } 
\label{fig:pmext}  
\end{figure}  
The point-mass approximation may not be appropriate for encounters where the binary star penetrates deeply within the perturber object, i.e. $b/r_h\simless 1$. On the other hand, in \S\ref{sec:catdif} we argue that collisions between dark matter substructures and stellar binary systems in the dark matter haloes of dSphs are expected to enter within the {\it catastrophic regime}. In this regime binary stars may be completely shattered by {\it single} encounters with dark substructures if $b\simless b_{\rm min}$, where $b_{\rm min}$ is defined in eq.~(\ref{eq:bmin}). Thus, fatal encounters with point-mass perturbers will occur within an interval of impact parameters $r_h < b <b_{\rm min}$. 

In the case of large, extended perturbers, however, we may find situtations where $r_h\simgreat b_{\rm min}$. The goal of this Appendix is to examine how the point-mass approximation may affect our results.

Let us consider first how the mass and the size of dark structures are related in CDM. Given that the mass and circular velocity of CDM haloes scale as $M\propto V^3$, and that the virial theorem is, $V^2\propto M/R$, it follows that the half-mass radius and the mass of dark substructures scale as $r_h\propto M_p^{1/3}$.
From eq.~(\ref{eq:bmin}) we find $b_{\rm min}/r_h\propto M_p^{1/2} a^{3/4}/r_h\sim M_p^{1/6} a^{3/4}$, which suggests that whether or not the point-mass approximation is valid is mainly set by the binary separation.

Scaling eq.~(\ref{eq:bmin}) to the parameters typically found in dSphs (see \S\ref{sec:catdif}) we find

\begin{eqnarray}
\frac{b_{\rm min}}{a}\approx 2\cdot 10^3 \bigg( \frac{M_p}{10^8 M_\odot} \frac{\sqrt{2} \cdot 10 {\rm km s}^{-1}} {V_{\rm rel}}  \bigg )^{1/2} \bigg(\frac {1 M_\odot}{M_b} \frac{0.1 {\rm pc}}{a}\bigg)^{1/4}.
\label{eq:bmin_esc}
\end{eqnarray}

Thus, for $b_{\rm min}/a\sim 2\cdot 10^3$ we expect the point-mass approximation to hold for binaries with very wide separations, $a\simgreat \big(b_{\rm min}/a\big)^{-1}r_h\sim 5\cdot 10^{-4}r_h$. Conversely, if we adopt a fiducial separation $a=0.1$ pc, the point-mass approximation will be valid for perturbers with half-mass radii $r_h\simless 200$ pc.

Unfortunately, there is no clear CDM prediction about the size of sub$^2$haloes orbiting in dwarf-size haloes, simply because no cosmological N-body simulation has yet reached the resolution required to study the internal structure of these systems in detail. Here we shall assume that the size-mass relation found for subhaloes (Diemand et al. 2008; Springel et al. 2008) also applies to sub$^2$haloes, and that both can be approximated by

\begin{equation}
r_h\approx 10^3 {\rm pc}\bigg( \frac{M_p}{10^9 M_\odot}\bigg)^{1/3} .
\label{eq:rhm}  
\end{equation}

 \begin{figure}  
\includegraphics[width=84mm]{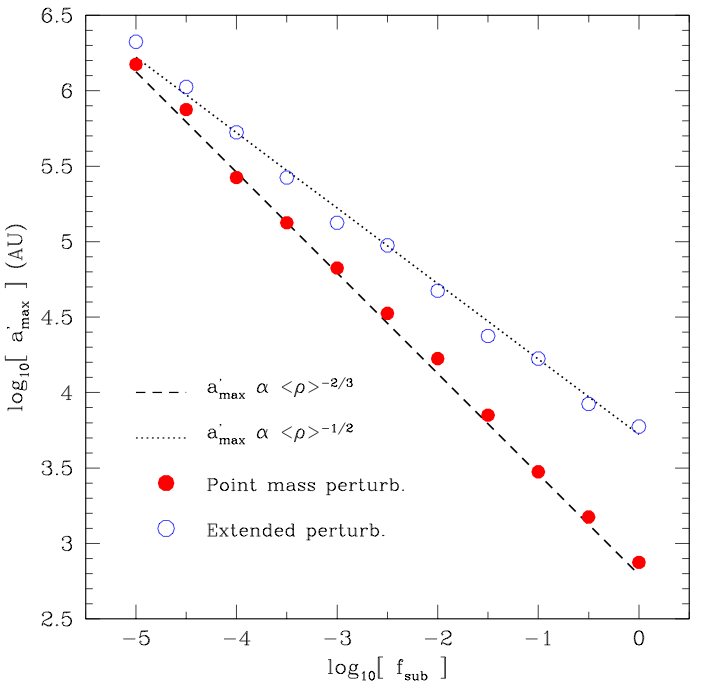}
\caption{Separation at which the binary distribution truncates as a result of encounters with point-mass (closed symbols) and extended (open symbols) substructures that amount a total mass of $f_{\rm sub}M_{\rm vir}$, where $M_{\rm vir}$ is the virial mass of the dwarf galaxy. The value of $a^{'}_{\rm max}$ is defined as the separation at which $d N_b/d\log_{10} a$ drops 20\% with respect to the unperturbed distribution. Dashed lines show the expected position of the truncation as computed from eq.~(\ref{eq:amax}). The dotted line is a fit to the values of $a^{'}_{\rm max}$ derived from our N-body simulations.}
\label{fig:amax2}  
\end{figure}  

In the upper and lower panels of Fig.~\ref{fig:pmext} we show the results of repeating the simulations presented in \S\ref{sec:test} adopting point-mass and extended sub$^2$halo models, respectively. In practice, the only difference between both calculations corresponds to the value of $U(b/r_h)$ adopted in eq.~(\ref{eq:de}), which is $U=1$ for point-mass perturbers, and $U(x)=x^{2.5}/(1+x)^{2.5}$ for extended ones (see \S8.4 of Binney \& Tremaine 2008).  

As already shown in \S\ref{sec:test}, if dark substructures are assumed to be point-mass objects eq.~(\ref{eq:amax}) provides an accurate estimate of the binary separation at which deviations from the unperturbed binary distribution, $d N_b/d a\propto a^{-1}$, start to become evident. At a separation $a=a_{\rm max}$ we find that the number of bound binaries drops 20\% with respect to the unperturbed distribution independently of the perturber density. 

Interestingly, if dark substructures are modelled as extended objects the truncated binary population also scales as $dN_b/d a \propto a^{-2.1}$ for $a\gg a'_{\rm max}$, where we define $a'_{\rm max}$ as the separation at which the number of binaries drops by 20\%. However, in this case eq.~(\ref{eq:amax}) underestimates the location of the truncation of the binary distribution, so that $a'_{\rm max}\simgreat a_{\rm max}$. 

Fig.~\ref{fig:amax2} shows how the maximum separation, $a'_{\rm max}$, varies as a function of the mean sub$^2$halo density $\langle\rho_p\rangle=f_{\rm sub}M_{\rm vir}3/(4\pi R_b^3)$ for point-mass (closed symbols) and extended (open symbols) bodies. It is clear that the point-mass approximation provides an accurate estimate of $a'_{\rm max}$ if the fraction of sub$^2$haloes represent a small fraction of the dwarf galaxy mass. This is because the number of massive substructures, which also tend to have larger half-light radii, is proportional to $f_{\rm sub}$. For $f_{\rm sub}\sim 1$, however, the point-mass approximation under-estimates $a'_{\rm max}$ by a factor $\approx 10$. 

Interestingly, Fig.~\ref{fig:amax2} also show that $a'_{\rm max}$ also scales as a power-law if perturbers are modelled as extended objects. In particular we find that $a'_{\rm max}\propto \langle \rho_p \rangle^{-1/2}$, a result that is used in \S\ref{sec:dSphs} to estimate the truncation in the separation function of the binary population orbiting in Milky Way dSphs.

\end{document}